\documentclass[
    ,final            
  ]
  {aipproc}

\layoutstyle{8x11double}

\begin{document}

\title{Simulated synchrotron and Inverse Compton emission from Pulsar Wind Nebulae}


\author{Delia Volpi}{
  address={Dipartimento di Astronomia e Scienza dello Spazio-Largo Fermi 2-50125 Firenze-Italy}
}

\author{Luca Del Zanna}{
  address={Dipartimento di Astronomia e Scienza dello Spazio-Largo Fermi 2-50125 Firenze-Italy}
}

\author{Elena Amato}{
  address={INAF-Osservatorio Astrofisico di Arcetri-Largo Fermi 5-50125 Firenze-Italy} 
}

\author{Niccol\'o Bucciantini}{
  address={Astronomy Department-University of California at Berkeley-601 Campbell Hall-Berkeley-CA 94720-3411-USA}
}

\begin{abstract}
We present a complete set of diagnostic tools aimed at reproducing synthetic non-thermal (synchrotron and/or Inverse Compton, IC) emissivity, integrated flux energy, polarization and spectral index simulated maps in comparison to observations. The time dependent relativistic magnetohydrodynamic (RMHD) equations are solved with a shock capturing code together with the evolution of the maximum particles energy. Applications to Pulsar Wind Nebulae (PWNe) are shown.  

\end{abstract}


\maketitle


Pulsar Wind Nebulae (e.g. Vela and Crab Nebula) are a class of Supernova Remnants (SNR), which originates from the interaction of the pulsar wind with the surrounding ejecta. They are characterized by non-thermal radiation at all wavelengths, mostly synchrotron (from radio to X-ray bands) and Inverse Compton (gamma-ray band), due to the presence of high energy pairs embedded in a strong magnetic field. In the optical and X-ray ranges observations from space (HST and Chandra) show typical axisymmetric features known as ``jet-torus structure''. 

The theoretical interpretation of the jet-torus morphology  
\cite{Bogovalov:2002}, \cite{Lyubarsky:2002} is based on the idea of a stronger  equatorial energy flux. This creates the torus and an oblate termination shock (TS) with cusps at the poles. Post-shock flows first converge toward the equator and then are diverted along the symmetry axis by magnetic hoop-stresses to form the jets. Thanks to the development of RMHD codes (i.e.  \cite{Del Zanna:2003}), we can now investigate this picture in detail. 

Our scheme \cite{Del Zanna:2004} evolves the ideal axisymmetric RMHD equations together with the local maximum particle energy (in unit of $\mathrm{m} \mathrm{c}^{2}$) $\epsilon_{\infty}$, taking in account of adiabatic and synchrotron losses:
\begin{equation}
\frac{\mathrm{d}\ln\epsilon_{\infty}}{\mathrm{dt}^{'}}=\frac{\mathrm{d}\ln \mathrm{n}^{1/3}}{\mathrm{dt}^{'}}+\frac{1}{\epsilon_{\infty}}\left(\frac{d\epsilon_{\infty}}{\mathrm{dt}^{'}}\right)_{\mathrm{sync}} \label{energy}           
\end{equation}
where the first term represents the adiabatic losses, the second one represents the synchrotron losses averaged with pitch angles, $\left(\frac{d\epsilon_{\infty}}{\mathrm{dt}^{'}}\right)_{\mathrm{sync}}=-\frac{4\mathrm{e}^{4}}{9\mathrm{m}^{3}\mathrm{c}{5}}\mathrm{B}^{' 2}\epsilon_{\infty}^{2}$, $\mathrm{n}$ is the fluid proper density, $\mathrm{m}$ and $\mathrm{e}$ are respectively the particle mass and charge, $\mathrm{c}$ is the velocity of light, $\mathrm{t^{'}}$ is the time and $\mathrm{B}^{'}$ is the amplitude of the magnetic field in the comoving frame. 

We employ the ECHO (Eulerian Conservative High Order Scheme) code developed by Del Zanna's code  \cite{Del Zanna:2003} (recently extended to GRMHD  \cite{Del Zanna:2007}) assuming a purely toroidal magnetic field and  poloidal velocity. A description of the initial and boundary conditions can be found in \cite{Del Zanna:2004} and \cite{Del Zanna:2006}.

Emitting particles are continuously injected at TS with a distribution function which is a power-law in energy and isotropic in momentum:
\begin{equation}
 \mathrm{f}(\epsilon_{\mathrm{TS}})=\frac{\mathrm{A}_{\mathrm{w}}}{4\pi}\epsilon_{\mathrm{TS}}^{-(2\alpha+1)},
 \end{equation}
where $\mathrm{A}_{\mathrm{W}}$ is proportional to the thermal pressure of the post-shock fluid (the electron thermal energy), $\epsilon_{\mathrm{TS}}$ is the particle energy at TS and $\alpha=0.7$ is the spectral index. The particles are then advected by the nebular mildly relativistic flow and the post-shock distribution function is obtained from conservation of particles'number along streamlines as in \cite{Del Zanna:2006}. 

Under the assumption of quasi-stationarity and of negligible synchrotron losses, the emission coefficient in observer's fixed frame is:
\begin{equation}
\mathrm{j}_{\nu}(\nu,\vec{\mathrm{n}}) \propto\ \mathrm{D}^{\alpha+2}\cdot \mathrm{p} \cdot |\vec{\mathrm{B}{'}}\times\vec{\mathrm{n}^{'}}|^{\alpha+1}\cdot \nu^{-\alpha}                   
\end{equation}
if $\nu_{\infty} \ge \nu$ and $0$ elsewhere. $\mathrm{D}$ is the Doppler factor, $\vec{\mathrm{n}^{'}}$ is the direction of the observer in the comoving frame, $\nu$ is the optical or X-ray observation frequency, $\nu_{\infty}$ is the cut-off frequency (used to obtain the synchrotron burn-off):
\begin{equation}
\nu_{\infty}=\mathrm{D}\frac{3\mathrm{e}}{4\pi \mathrm{m c}}|\vec{\mathrm{B}{'}}\times\vec{\mathrm{n}^{'}}|\epsilon^{2}_{\infty}.
 \end{equation}

Integrating the emission coefficient along the line of sight one can obtain surface brightness, optical polarization \cite{Bucciantini:2005} and spectral index maps \cite{Del Zanna:2006}.

\begin{figure}
\label{figures1}
  \includegraphics[height=.2\textheight]{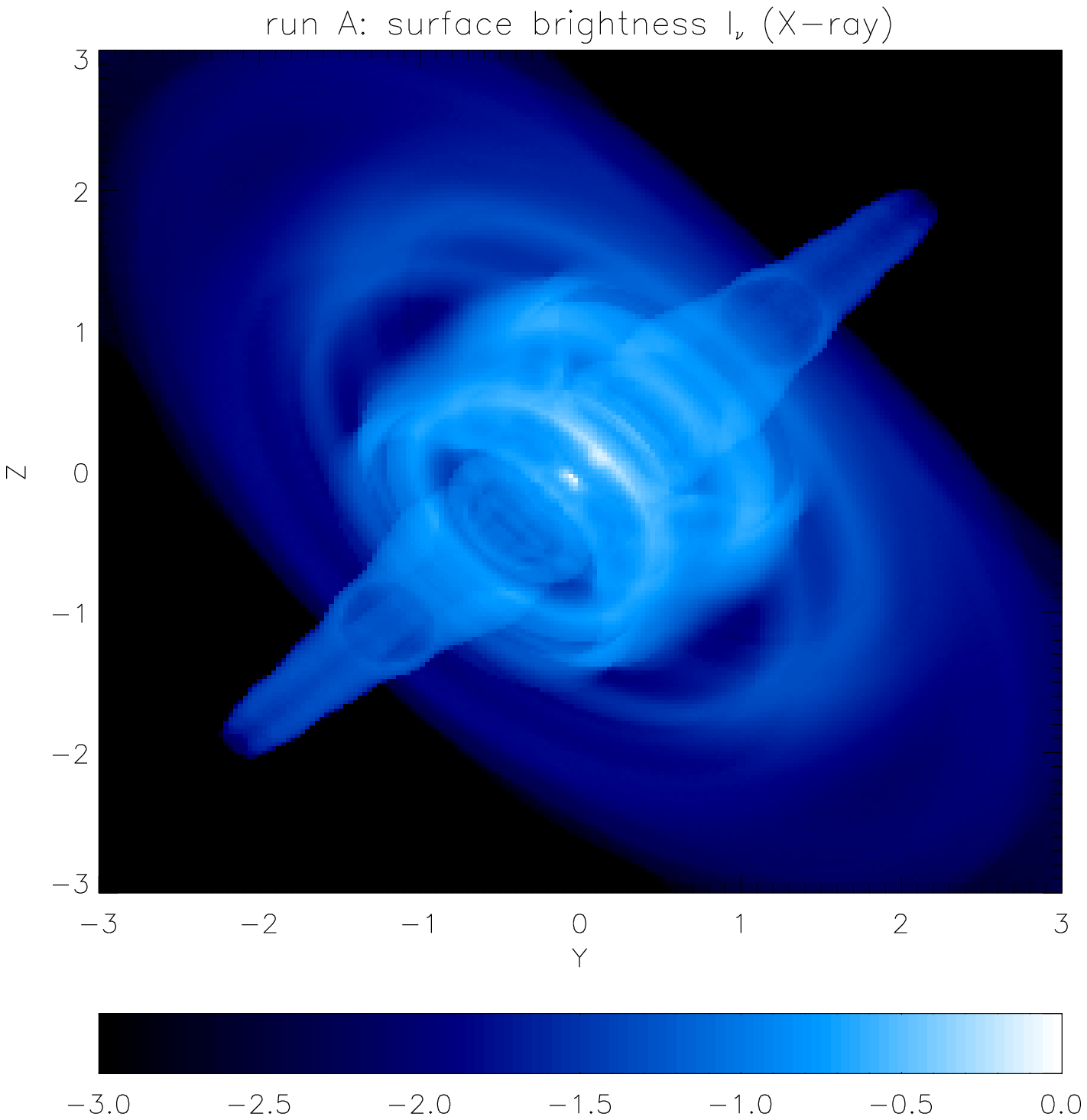}
\includegraphics[height=.2\textheight]{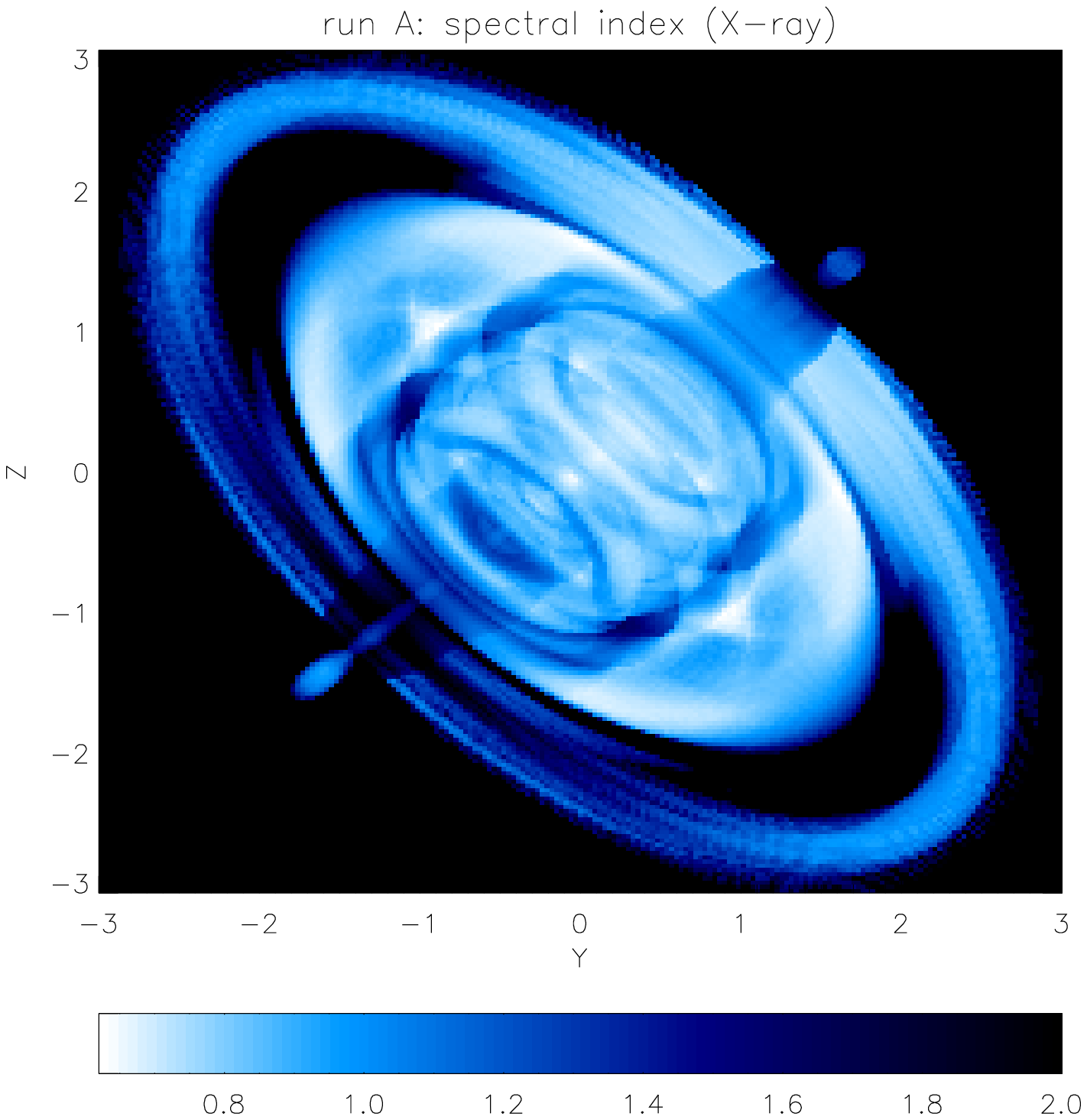}
  \caption{On the left: runA ($\mathrm{b}=10$ and $\sigma=0.025$) simulated brightness map in X-ray band, in logarithmic scale and normalized to the maximum value. On the right: runA simulated spectral index map in X-ray band. Distances from the central pulsar are reported on the axes, expressed in light year (ly) units.}
\end{figure}

Simulated flux maps reproduce the two polar jets and the equatorial flows with the correct ranges of velocity. Brightness maps show, as expected, an equatorial torus, a system of rings, with brighter arches and a central knot due to Doppler boosting (Fig.\ref{figures1}). The optical and X-ray spectral index maps (Fig.\ref{figures1}) agree with observations \cite{Veron-Cetty:1993}, \cite{Mori:2004}.

We found that, by changing the width of the equatorial striped region $\mathrm{b}$ and the magnetization of the wind $\sigma$ (with $\sigma_{\mathrm{effective}}>0.1$ to have supersonic polar jets), the emission pattern changes so that it is, in principle, possible to use the observed morphology of the jet-torus structure to infer the conditions in the wind and to reproduce different Pulsar Wind Nebulae.

High energy gamma-ray emission (COMPTEL, EGRET, HEGRA and HESS observations) is primarily due to IC scattering of high energy electrons on target photons from CMB, dust (far infra-red emission, FIR), starlight, and from the PWN itself (synchrotron emission, SSC). 

We consider here two different post-shock distribution functions \cite{Atoyan:1996}: one describing the primordial radio emitting electrons (born at the supernova outburst) and the other the high energy tail accelerated at the TS. Our aim is to reproduce the observed spectrum in Crab.

The low energy distribution function is:
\begin{equation}
\mathrm{f}_{\mathrm{r}}(\epsilon)=\mathrm{A}_{\mathrm{r}}\epsilon^{-(2\alpha_{\mathrm{r}}+1)}e^{-\epsilon/\epsilon_{*}}
 \end{equation}
with $\mathrm{A}_{\mathrm{r}}$ a constant chosen to fit data at $1$Ghz, $\alpha_{r}=0.26$ the spectral index and $\epsilon_{*}$ the radio particle energy cut-off.

The high energy distribution function at TS is:
\begin{equation}
\mathrm{f}_{\mathrm{TS}}(\epsilon_{\mathrm{TS}})=\mathrm{A}_{\mathrm{w}}(\epsilon_{\mathrm{TS}}+\epsilon_{0})^{-(2\alpha_{\mathrm{w}}+1)}e^{-\epsilon_{\mathrm{TS}}/\epsilon_{\mathrm{c}}}
 \end{equation}
where $\mathrm{A}_{\mathrm{w}}$ is a constant proportional to the thermal pressure of the post-shock fluid, $\epsilon_{0}$ is the lowest wind particle energy chosen to fit radio-optical data,
$\alpha_{\mathrm{w}}=0.7$ is the spectral index and $\epsilon_{\mathrm{c}}$ is the gamma particle energy cut-off at MeV frequencies.

The incident photon density per unit of frequency is obtained either from a black body formula (CMB or FIR target) or from integrated synchrotron flux (SSC) assumed to be homogeneous and isotropic.

Integrating distribution function and power per unit of frequency over the particle energy and incident photon frequencies for every energy regime \cite{Blumenthal:1970}, energy spectra and brightness maps are produced. 

\begin{figure}
\label{figures2}
  \includegraphics[height=.2\textheight]{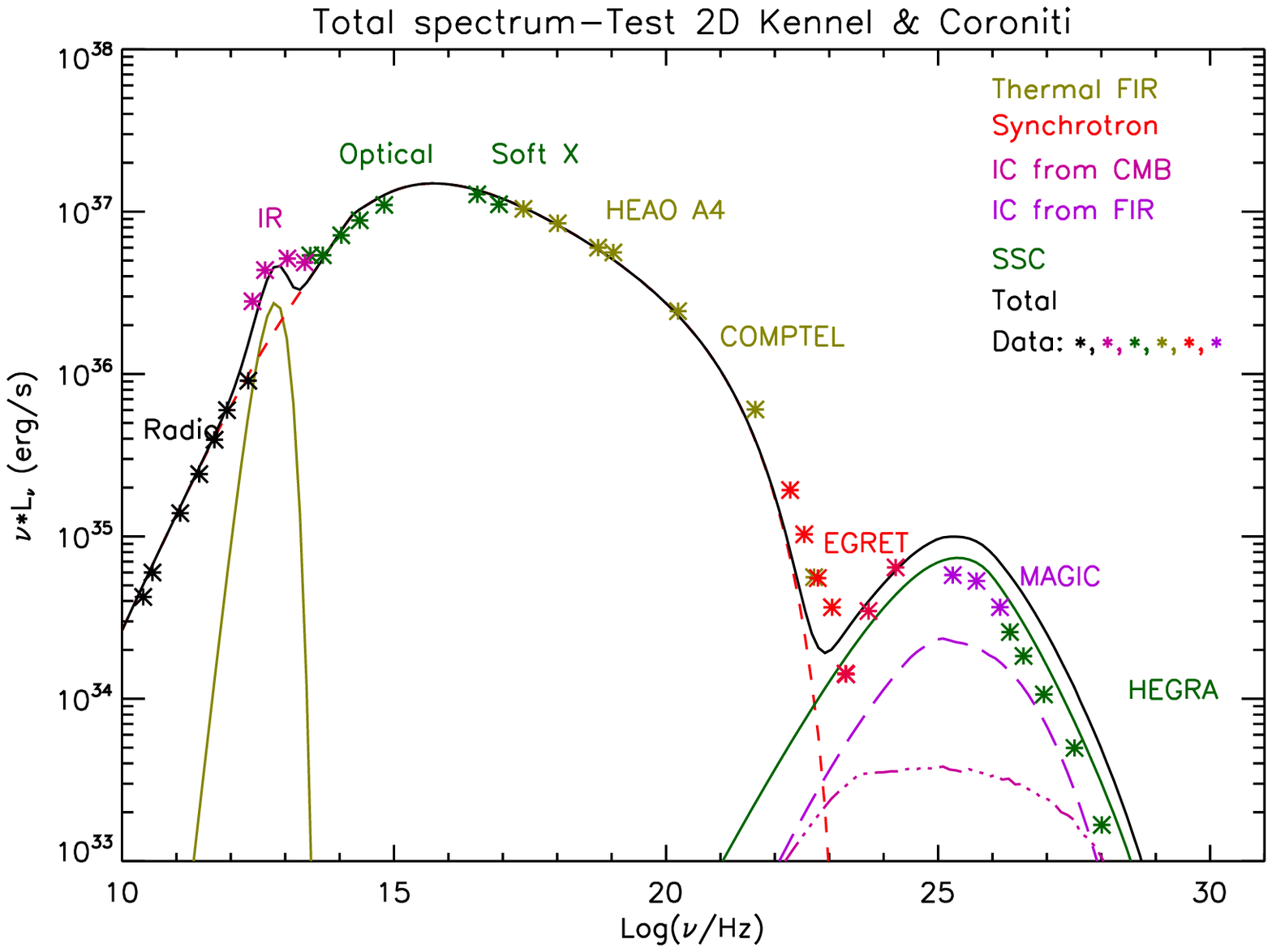}
\includegraphics[height=.2\textheight]{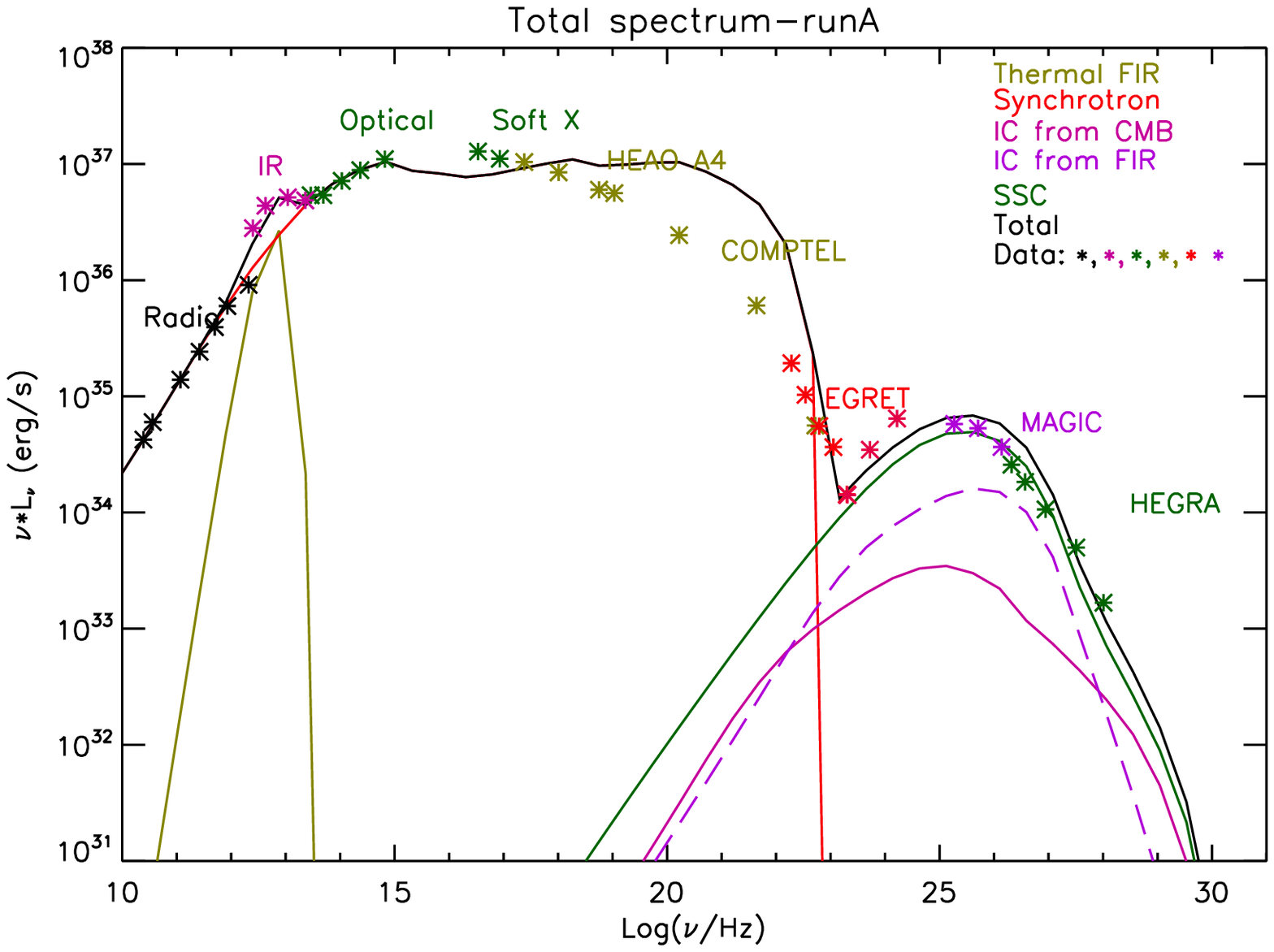}
  \caption{On the left: Kennel and Coroniti 2-D spectra. On the right: runA spectra. Crab observed data are also plotted.}
\end{figure}

\begin{figure}
\label{figures3}
  \includegraphics[height=.2\textheight]{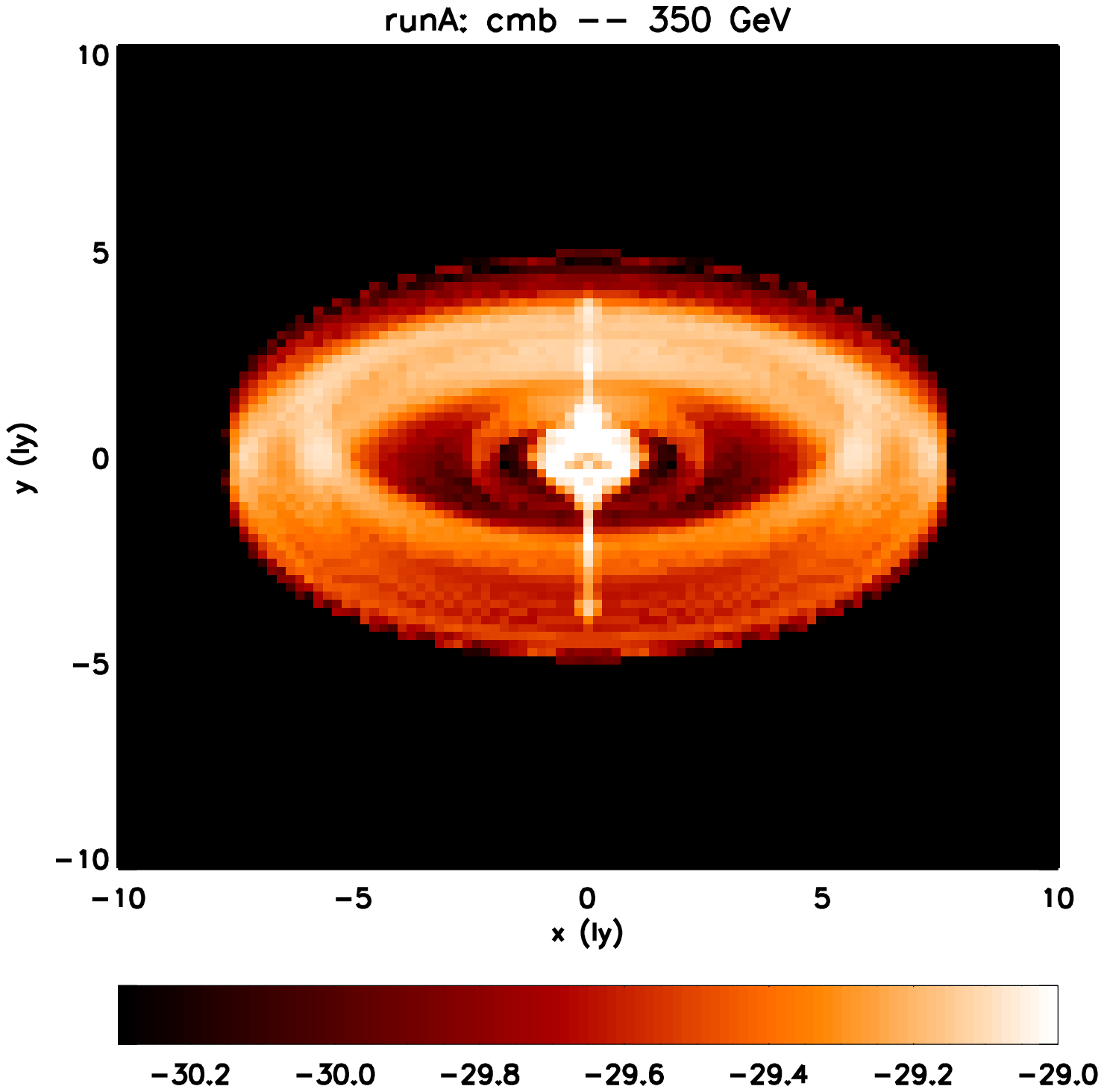}
\includegraphics[height=.2\textheight]{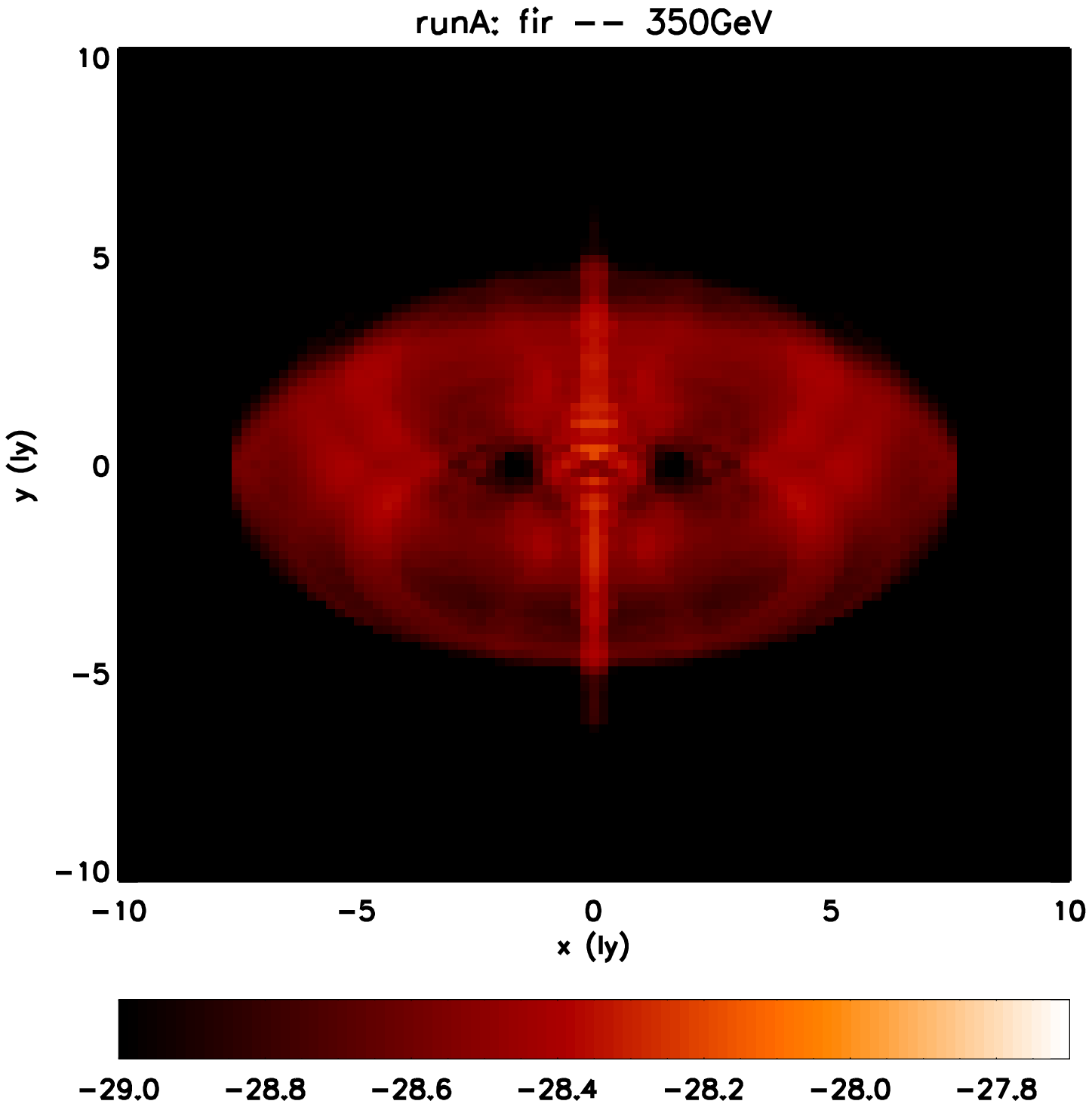}
\includegraphics[height=.2\textheight]{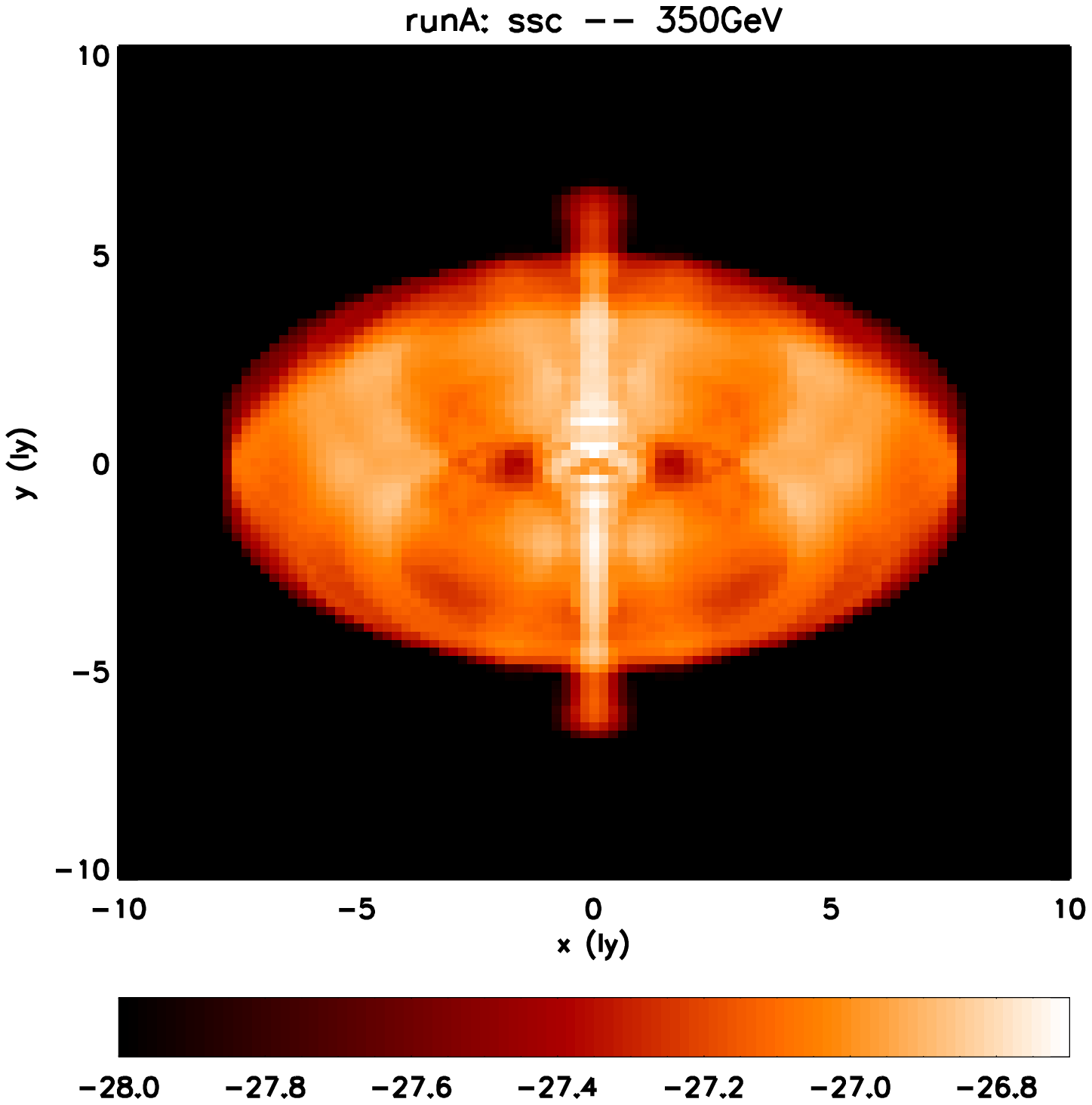}
  \caption{Simulated brightness maps (in units of $\mathrm{erg}/(\mathrm{sr}\; \mathrm{cm}^{2}\mathrm{s}\; \mathrm{Hz}$) in logarithmic scale) at 350GeV. On the left: IC from CMB target. In the middle: IC from FIR target. On the right: IC from synchrotron target.}
\end{figure}

The emission recipes has been successfully tested by comparing the broadband 
spectrum obtained with the Kennel and Coroniti spherical flow \cite{Kennel:1984}, \cite{Atoyan:1996} against the Crab Nebula observations (fig. \ref{figures2}). However, when applied to the flow structure from RMHD simulations, results at synchrotron high energies are not satisfactory yet and further tuning of the emission model would be required to fit all data (Fig.\ref{figures2}). Gamma-ray surface brightness maps are shown in Fig.\ref{figures3} for various sources of target photons. The strongest IC emission is due to SSC. 

The present work confirms jet-launching mechanism due to magnetic 
hoop stresses with the best agreement between simulations and Crab Nebula data given by the wind magnetization parameter $\sigma_{\mathrm{effective}} \approx 0.02$.

Further work is required to refine our model in order to compare with future Glast observations (paper in preparation).

This complete set for calculating simulated synchrotron emission, polarization 
and spectral index maps accounting for synchrotron losses can be used for other classes of objects (eg. AGN jets), in any scheme for RMHD (e.g. non-conservative, in full 3-D).


\begin{theacknowledgments}
N.B. was supported by NASA through Hubble Fellowship grantHST-HF-01193.01-A, awarded by the Space Telescope Science Institute,which is operated by the Association of Universities for Research inAstronomy, Inc., for NASA, under contract NAS 5-26555.
\end{theacknowledgments}

\bibliographystyle{aipproc}   


\bibliography{sample}

\end{document}